\documentclass[twocolumn,aps,prb,amsmath,showpacs]{revtex4}
\usepackage{graphicx,dcolumn,bm,amssymb,amsmath}
\usepackage{color}

\def\be{\begin{equation}}
\def\ee{\end{equation}}

\begin{document}

\title{Network disorder and non-affine deformations in marginal solids}
\author{Alessio Zaccone, Jamie R. Blundell and Eugene M. Terentjev}
\affiliation{Cavendish Laboratory, University of Cambridge, JJ Thomson Avenue,
Cambridge CB3 0HE, U.K.}
\date{\today}
\begin{abstract}
\noindent The most profound effect of disorder on the elastic response of
solids is the non-affinity of local displacements whereby the atoms (particles,
network junctions) do not simply follow the macroscopic strain, as they do in
perfect crystals, but undergo additional displacements which result in a
softening of response. Whether disorder can produce further effects has been an
open and difficult question due to our poor understanding of non-affinity. Here
we present a systematic analysis of this problem by allowing both network
disorder and lattice coordination to vary continuously under account of
non-affinity. In one of its limits, our theory, supported by numerical
simulations, shows that in lattices close to the limit of mechanical stability
the elastic response stiffens proportionally to the degree of disorder. {This
result has important implications in a variety of areas: from understanding the
glass transition problem to the mechanics of biological networks such as the
cytoskeleton.}
\end{abstract}
\maketitle

\section{\textbf{Introduction}}
In a disordered lattice subject to a small external deformation, every atom or
network junction moves initially in an affine way, that is, displaced in
geometric proportion to the macroscopic strain. At the same time, the atom
experiences forces from its nearest-neighbors (NN) which are initially
displaced affinely too. In any disordered lattice atoms will therefore
experience a non-zero force due to the fact that their NN are randomly
oriented. By contrast, in a perfectly ordered lattice the forces transmitted by
the NN may balance each other by symmetry and the resultant local force in that
case can be zero. The finite forces acting on each atom in the disordered
network drive additional local displacements, called non-affine
displacements~\cite{alexander,lubensky,lemaitre}.  As these are accompanied by
a release of energy, needed in order to reach mechanical equilibrium on each
atom, the effect of non-affinity induced by the disorder on the macroscopic
rigidity is always to soften the solid with respect to a purely affine
deformation. This softening is a well established phenomenon observed many
times in simulations and experimental studies~\cite{barrat,behringer}. However,
since a consistent theoretical description of non-affine elastic response has
been elusive so far, it has been difficult to predict the rigidity of
disordered solids (including glass, granular media, cytoskeletal networks,
etc.) as a function of the degree of structural order/disorder. Apart from its
obvious importance for materials science, rigidity is the defining property of
the solid state of matter~\cite{anderson} and furthermore the concept of
(generalized) rigidity as the epitome of cooperative behavior is essential for
various other properties, including how information is transmitted in many-body
systems~\cite{anderson_book}. In spite of this, the relation between rigidity
and disorder has remained largely unknown~\cite{vanhecke}. This may well be the
reason for our present lack of understanding of the glass transition
phenomenon, wherein rigidity emerges in a (disordered) supercooled liquid
without any apparent change in the symmetry of disordered
structure~\cite{anderson_book,debenedetti}.

Here we report on a systematic theoretical and numerical study of the elastic
rigidity (described through the shear modulus) as a function of the degree of
structural disorder, which varies from the perfect crystal all the way up to
the completely disordered lattice. We focus our analysis to central-force
systems~\cite{thorpePRB} where the physical bonds between particles have zero
bending rigidity, {and to systems which are close to the vanishing of rigidity
(\emph{marginal solids})}. A wide class of physical systems are captured by these
central force models, such as flexibly-linked biological networks and metallic
glasses. In some cases, such as covalently bonded materials, junctions can
possess bending rigidity which makes the transition to a rigid phase occur at
lower connectivity~\cite{thorpePRL,mackintosh,frey}. Without loss of
generality, our results provide the first direct evidence for the existence of
a broad range of structural disorder where increasing the disorder causes
lattices near the isostatic limit {(i.e. the limit where the number of
geometric constraints exactly matches the number of degrees of freedom)} to
stiffen. This result has deep implications in view of the fact that all liquids
on their way toward the solid state do pass through this isostatic limit,
whereby even a small difference of mechanical stability with respect to
internal stresses generated by vitrification can decide their ultimate fate,
i.e. whether to become permanently ordered (crystal) or fully disordered
(glass).

\section{\textbf{The Gaussian bond-orientational disorder model}}
Let us start by introducing the bond-orientation vectors between
nearest-neighbor atoms $i$ and $j$ defined by
$\underline{n}_{ij} = (\cos {\phi}\sin {\theta},\sin {\phi}\sin {\theta},\cos
{\theta})$. The affine part of the shear modulus of harmonic lattices can be
derived from the expansion of the free energy in the harmonic approximation and
is given in this limit by the well-known Born-Huang formula~\cite{born}, for
instance,
$C_{xyxy}^A =(1/V)\kappa R_0^2\sum\nolimits_{ij} {{c_{ij}}}
n_{ij}^xn_{ij}^yn_{ij}^xn_{ij}^y$, where the lattice sum goes over all
nearest-neighbor pairs $ij$. The atomic force constant  $\kappa$  is defined by
the harmonic pair-potential, $U({r_{ij}}) = \frac{1}{2}\kappa {({r_{ij}} -
{R_0})^2}$, between nearest-neighbors. $R_{0}$ is the interatomic distance at
rest in the reference frame and $V$ is the volume of the system before
deformation. Here and below the Roman indices are used to label atoms while
Greek indices are used to label Cartesian components of vectors. $c_{ij}$ is
the connectivity matrix with $c_{ij} =1$ if $i$ and $j$ are nearest neighbors
and $c_{ij}=0$ otherwise in our simple case. Let us consider the shear modulus
$G$ of a simple cubic lattice in $d=3$ in the affine approximation. Every atom
has $z=6$ nearest neighbors and shares a bond with each of them. With $N$ atoms
in the lattice, on average each atom contributes $z\langle
n_{ij}^xn_{ij}^yn_{ij}^xn_{ij}^y\rangle$ to the lattice sum, where the
bond-orientation averaged factor is evaluated with an appropriate, normalized
orientation distribution function (ODF) $f(\theta ,\phi )$:
\begin{equation}
\langle n_{ij}^xn_{ij}^yn_{ij}^xn_{ij}^y\rangle  = \int {f(\theta ,\phi )}
{\sin ^4}\theta {\cos ^2}\phi {\sin ^2}\phi \sin \theta d\theta d\phi
\label{eq:nnnn}
\end{equation}
Using this averaging the Born-Huang formula can be written:
\begin{equation}
C_{xyxy}^A =(1/V)\kappa R_0^2 z\langle n_{ij}^xn_{ij}^yn_{ij}^xn_{ij}^y\rangle.
\label{eq:born}
\end{equation}
Evaluating $f(\theta ,\phi )$ for different physical situations is one of the
main exercises of our theory, leading to different forms of the resulting shear
modulus. For the simple cubic system, out of the six orientations of the bonds
attached to each atom, only two are independent, while the other $4$ can be
obtained from these by symmetry operations that leave products of
$n_{ij}^\alpha$ terms ($\alpha  = {\rm{x}},{\rm{y}},{\rm{z}}$ ) unchanged.
These symmetry operations are either rotations (belonging to the octahedral
rotation symmetry group $O_{h}$) about the zenithal (z-)axis, or reflections
through any plane of symmetry of the cubic cell. One of the two independent
orientations lies along the zenithal axis and is parameterized by
${\theta_{ze}} = 0$ and $\forall {\phi_{ze}}$, while the other one lies in the
azimuthal plane and is parameterized by ${\theta_{az}} = \pi /2$ and
${\phi_{az}} = 0$ (see Fig.1a).
 The former occurs with probability 2/6 and the latter one with probability
4/6. Hence the normalized ODF of the simple cubic (SC) lattice can be
rewritten, for our purposes, in a simpler normalized form which is going to be
used throughout (see the Appendix A for full details):
${f_{SC}}(\theta ,\phi ) = (3/2)[(2/6)\delta (\theta ) + (4/6)\delta (\theta  -
\pi /2)\delta (\phi )]$.  Using this ODF in Eq.(\ref{eq:nnnn}) immediately
leads to ${\langle n_{ij}^xn_{ij}^yn_{ij}^xn_{ij}^y\rangle _{SC}} = 0$ and so
the affine shear modulus for the SC lattice is ${G^A} = C_{xyxy}^A
=(N/V)z\kappa R_0^2{\langle n_{ij}^xn_{ij}^yn_{ij}^xn_{ij}^y\rangle _{SC}} =
0$. This result given by the affine theory for the perfectly ordered simple
cubic lattice agrees with Maxwell's counting of degrees of
freedom~\cite{thorpe_phillips} which indeed predicts $G=0$ for isostatic
lattices with $z=2d=6$, and also with evidence from polonium-210 in its alpha
form, and the superconducting calcium-III, which both feature simple cubic
structure and a vanishing shear stability at ambient
pressure~\cite{yabuuchi,djohari,legut}.

\begin{figure}
\includegraphics[width=0.8\linewidth]{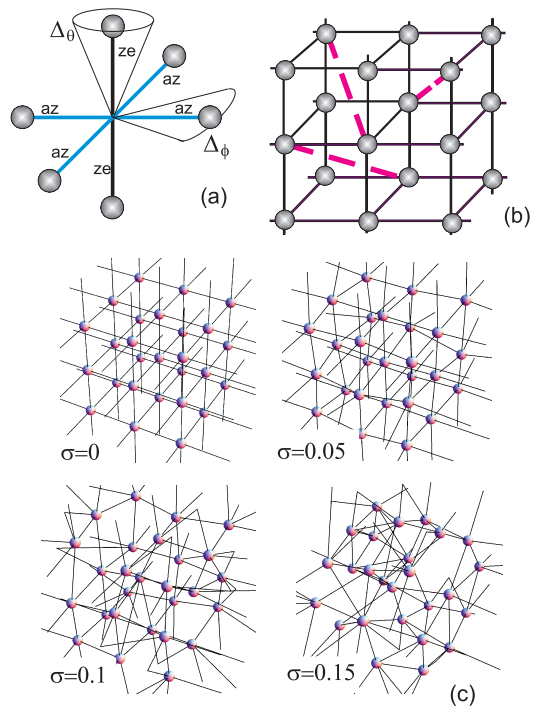}
\caption{(a) Primitive cell of the (perfectly ordered) simple cubic lattice.
The bonds lying on the horizontal plane are called azimuthal bonds (az); those
on the vertical axis are zenithal (ze). (b) Schematic model of cubic lattice
with randomly added extra bonds (dashed lines). (c) Marginal solids with
varying degree of disorder parameterized by the degree of disorder $\sigma$
defined in the text.}
\label{fig1}
\end{figure}

In the opposite limit of a completely disordered network of harmonic springs,
one may assume that all bond orientations occur with the same probability. This
results in a completely de-correlated (bond-orientational) disorder described
by the uniform ODF ${f_{UD}}(\theta ,\phi ) = 1/4\pi $
which leads to ${\langle n_{ij}^xn_{ij}^yn_{ij}^xn_{ij}^y\rangle _{UD}} =
1/15$, where we use the subscript UD for ``uniform disorder''. Hence using this
result in Eq.(\ref{eq:born}) gives $G = C_{xyxy}^A = (1/30)(N/V)\kappa R_0^2z$,
which is evidently finite for any coordination number $z$. Thus the affine
approximation here dramatically fails to comply with the vanishing of rigidity
at isostaticity point (i.e. at $z=6$), in contrast with both Maxwell's
constraint-counting and the empirical evidence~\cite{vanhecke}. This represents
the well-documented failure of the affine approximation to describe the
elasticity of disordered lattices~\cite{vanhecke,zaccone,makse}.

Let us now introduce a way to quantify arbitrary degrees of structural
disorder. We again consider the simple cubic lattice as our starting point and
introduce disordered realizations, always keeping $z=6$ for the time being. The
orientations of the bonds now deviate from the cubic lattice axes, and we
assume that their distribution is Gaussian about these axes. A particularly
convenient choice to parameterize Gaussian bond-orientational disorder in the
network of this topology is the following ODF: ${f_{GD}} = (2/6)f_{GD}^{ze} +
(4/6)f_{GD}^{az}$, with:
\begin{equation}
\begin{aligned}
f_{GD}^{ze} &= {N_{ze}}({\sigma ^2}){e^{ - {\theta ^2}/2{\sigma ^2}}} \\
f_{GD}^{az} &= {N_{az}}({\sigma ^2}){e^{ - {{(\theta  - \frac{\pi
}{2})}^2}/2{\sigma ^2}}}{e^{ - {\phi ^2}/2{\sigma ^2}}}
\end{aligned}
\label{eq:fGD}
\end{equation}
where the subscript GD stands for ``Gaussian disorder'' and $\sigma^2$ is the
variance, which acts as our quantitative measure of lattice disorder.
$N_{ze}(\sigma^2)$ and $N_{az}(\sigma^2)$ are normalization factors which also
depend on $\sigma^2$. We have verified that using this compact form for the
ODF, in terms of the two independent directions only, yields quantitatively the
same results that one would obtain by considering explicitly all the 6 bond
orientations.
This ODF allows one to span all intermediate degrees of disorder by varying the
parameter $\sigma^2$. In the limit ${\sigma ^{\rm{2}}} \to \infty $
this ODF flattens and one recovers the uniform orientational distribution
${f_{GD}}(\theta ,\phi ) \to {f_{UD}}(\theta ,\phi ) = 1/4\pi$. In the opposite
limit of ${\sigma ^{\rm{2}}} \to 0$, delta-functions are developed at the
angles of the cubic lattice so that one recovers the simple-cubic ODF,
${f_{GD}}(\theta ,\phi ) \to {f_{SC}}(\theta ,\phi )$.

\section{\textbf{Analytical theory of non-affine elastic deformations}}
\subsection{\textbf{Simple cubic and uniformly disordered lattices}}
The ultimate cause of non-affinity lies in the fact that the particles bonded
to a tagged particle, upon the externally applied strain are initially
displaced affinely, and owing to their displacement they exert a net force on
the tagged particle. If the particles were placed in an ordered fashion around
the tagged particle, such as in crystals, the resultant sum of these forces
would be zero due to symmetry. In a disordered solid such a cancelation does
not occur. Hence the resultant force is finite, and it induces an additional
displacement on the tagged particle in order to keep the mechanical
equilibrium, which adds to the affine displacement dictated by the imposed
strain. In this way the mechanical equilibrium is preserved on all particles.
Clearly, there is a work associated with these non-affine displacements, which
bears a negative sign since the overall energy is reduced. Formally this work
represents an additional term in the Born free energy expansion and can be
written as
\begin{equation}
 - W = \sum\limits_i {\int_0^{{\underline{\mathbf{u}}^{NA}}} {{f_{i\alpha }}dx_i^\alpha } }
=  - \frac{1}{2}\sum\limits_{ij} {H_{ij}^{\alpha \beta }} {\left. {x_i^\alpha
x_j^\beta } \right|_{{\underline{\bf{x}}} = {{\underline{\mathbf{u}}}^{NA}}}}.
 \label{eq:work}
\end{equation}
The sum runs over all particles and the path integral is over the non-affine
displacements ${\underline{\mathbf{u}}}^{NA}$ which the particle $i$ undergoes under the
action of the force field $\underline{f}$ exerted by the neighbors.
$H_{ij}^{\alpha \beta } = {\partial ^2}U/\partial r_i^\alpha \partial r_j^\beta
$ is the Hessian matrix and $x_i^\alpha $ are the coordinates of particle $i$.
Then the free energy expansion for a disordered solid accounting for
non-affinity is: $\delta F = \delta {F_A} - W$ where $\delta {F_A}$ is the
affine (Born) part of the free energy expansion under a given
strain~\cite{born}, which leads to the Born-Huang formula for the elastic
moduli already used above.

The negative non-affine contribution can be handled analytically (further
technical details can be found in the Appendix A). It has been recently shown
by Lemaitre and Maloney by normal mode decomposition~\cite{lemaitre} that the
non-affine contribution to the modulus takes the form
\begin{equation}
C_{\iota \xi \kappa \chi }^{NA} = \frac{1}{V}\sum\limits_k^{3N}
{\frac{{({{\underline \Xi  }_{\iota \xi}} \cdot {{\underline {\rm{v}}
}_k}{\rm{)}}({{\underline \Xi  }_{\kappa \chi}} \cdot {{\underline {\rm{v}}
}_k}{\rm{)}}}}{{{\lambda _k}}}},
 \label{eq:Cna}
 \end{equation}
 where ${\underline{\Xi}  }_{\iota \xi }$ are the affine (force) fields acting
on each particle, while ${\underline {\rm{v}}}_k$ and ${\lambda _k}$  are the
eigenvectors (which coincide with the normal modes of the solid) and
eigenvalues of the Hessian matrix, respectively. The sum runs over all
translational modes (degrees of freedom) of the atoms. Both the affine force
fields and the eigenvectors ${\underline {\rm{v}}}_k$ depend on the orientations
$\underline{n}_{ij}$ of the bonds in the solid in a way geometrically similar
to the way the bonds react elastically (affinely) to the imposed shear
deformation. Accordingly, the non-affine term is also proportional to the same
$\langle n_{ij}^xn_{ij}^yn_{ij}^xn_{ij}^y\rangle $ factor that has been present
in the affine model discussed above. As shown in the Appendix A, an allowed and convenient form 
of the eigenmodes can be found which leads to the exact evaluation of the non-affine terms in the sum in Eq.(\ref{eq:Cna}).
Taking the average over the disorder of the terms in the sum leads to $G^{NA} \equiv
C_{xyxy}^{NA}$:
\begin{figure}
\includegraphics[width=0.85\linewidth]{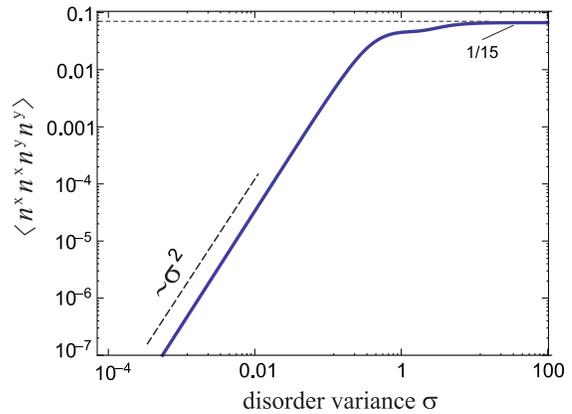}
\caption{The theoretical structure-dependent factor in the shear modulus,
plotted as a function of the degree of Gaussian disorder $\sigma$ calculated
using Eqs. (1) and (3). The asymptotic value of the strong-disorder plateau and
the scaling at weak disorder are shown by dashed lines. }
\label{fig2}
\end{figure}
\begin{equation}
{G^{NA}}  = (3N)\kappa R_0^2{\mkern 1mu} \langle
n_{ij}^xn_{ij}^yn_{ij}^xn_{ij}^y\rangle /V.
 \label{eq:Gna}
 \end{equation}
The important result is that the non-affine part of the shear modulus is in
fact proportional to the same angular average as the affine terms discussed
above (because the affine fields which are responsible for the non-affine term
have the same dependence on disorder as the affine part), and also to the total
number of degrees of freedom in the network, $3N$, which all take part in the
force relaxation. The non-affine effect is, therefore, independent of the
connectivity $z$ because the non-affine relaxation involves the degrees of
freedom rather than the bonds.

In the limit of uniform disorder we may recall that $G_{}^A = (1/30)(N/V)\kappa
R_0^2z$. Subtracting the corresponding uniform disorder-averaged expression for
the non-affine part,  $G_{}^{NA}$, we obtain the shear modulus as
\begin{equation}
G = \frac{{1}}{{{30}}}\frac{N}{V}\kappa R_0^2z -
\frac{{1}}{{{15}}}\frac{3N}{V}\kappa R_0^2{\mkern 1mu} =
\frac{{N}}{{{30V}}}\kappa R_0^2(z - 6).
 \label{eq:gUD}
 \end{equation}
This equation is derived in the Appendix A. This modulus correctly vanishes at
$z=6$, that is, at the isostatic point of central-force solids where the number
of constraints (bonds) equals the number of degrees of freedom. For $z < 6$
there is a finite number of floppy modes and the systems is no longer
rigid~\cite{thorpe_phillips}.

{Equation~(\ref{eq:gUD}) is based on the requirement that the Hessian averaged
over the orientational disorder be a $3N\times3N$ matrix where the $3\times3$
submatrices coincide with the identity matrix, as shown in the Appendix A. This
requirement is satisfied for both the simple cubic lattice and the uniformly
disordered lattice that have been defined above, i.e. for both the lower and
the upper limit of the Gaussian disorder spectrum that we introduced in the
previous section. Even though Eq.(\ref{eq:gUD}) may not hold exactly for
degrees of disorder intermediate between these two limits, we shall continue to
assume that Eq.(\ref{eq:gUD}) still applies as an approximate relation in between
the two limits. In the worst case this may give a reasonable interpolation over
the whole disorder spectrum. As this is a rather uncontrolled assumption, it
will be tested in comparison with numerical simulations later in this article.
}

Since the same orientation average $\langle
n_{ij}^xn_{ij}^yn_{ij}^xn_{ij}^y\rangle $ appears as a factor in both $G^{A}$
and $G^{NA}$, it follows that it plays the same role for an arbitrary degree of
disorder, that is, $G \propto {\langle n_{ij}^xn_{ij}^yn_{ij}^xn_{ij}^y\rangle
_{GD}}$, where the average encodes the information about the lattice disorder
via ${f_{GD}}(\theta ,\phi )$. The dependence of the microstructure function
${\langle n_{ij}^xn_{ij}^yn_{ij}^xn_{ij}^y\rangle _{GD}}$ upon the degree of
disorder as parameterized by $\sigma^2$ is explicitly shown in Fig.~2. In the
limit ${\sigma ^{\rm{2}}} \to \infty$, the completely uncorrelated uniform
disorder case is recovered, ${\langle n_{ij}^xn_{ij}^yn_{ij}^xn_{ij}^y\rangle
_{GD}} \to 1/15$. In the opposite limit of weak disorder, the structure
function goes to zero proportionally to $\sigma^2$, recovering the simple-cubic
lattice at $\sigma=0$. The deviation from this proportionality starts to be
significant when $\sigma^2\approx 1$, after which the uniform-disorder plateau ($=1/15$)
sets in.

\subsection{\textbf{Gaussian bond-orientational disorder with third-body
neighbors}}
We have therefore established that in the simplest situation, where all $z$ NN
atoms participate of the same degree of disorder, the effects of coordination
and disorder are decoupled and they contribute two distinct factors to the
shear modulus, $z-6$ (the coordination)  and ${\langle
n_{ij}^xn_{ij}^yn_{ij}^xn_{ij}^y\rangle _{GD}}$ (the disorder). An important
extension arises when bonds between further apart atoms (network junctions) are
added on a pre-existing lattice. This situation describes higher coordination
networks and occurs, for example, in cytoskeletal networks where filaments are
superimposed on an underlying network of a different type of filaments.
Clearly, the bond orientations of the added filaments will be  distributed
differently from the underlying network and thus have a different degree of
disorder. In order to capture this extension within our theory, we consider an
underlying isostatic network ($z=2d=6$) with a variable degree of disorder onto
which additional bonds are superimposed and placed at random at the network's
junctions. This scheme follows the most common model employed in the study of
\emph{marginal solids}~\cite{alexander,wyart,lubensky_2010} and consists of
taking an underlying ordered isostatic lattice and then randomly place
additional springs typically connecting next-nearest-neighbor or third-body
neighbor atoms (TBN).

Let us identify the bonds belonging to the underlying isostatic lattice as
isostatic bonds, while the additional bonds will be referred to as excess bonds
(see Fig.1b). On average, out of $z$ bonds per atom there are $6$ isostatic
bonds of a base cubic lattice, which we assume is formed from the variable
(Gaussian) disorder, and $z-6$  excess bonds per atom. Let us assume the excess
bonds are placed between TBN atoms along the diagonal of the cubic cell, so
their average length is of the order of $\sqrt 3 {R_0}$. The four third-body
neighbor excess bonds in the upper half-plane are defined by $\theta=\pi/4,
\forall \phi$, and $\phi=\pi/4, 3\pi/4, 5\pi/4, 7\pi/4$  while the four TBN
bonds in the lower half-plane are parameterized by $\theta=\pi/4+\pi/2, \forall
\phi$, and $\phi=\pi/4, 3\pi/4, 5\pi/4, 7\pi/4$. Hence, on average a randomly
placed third-body neighbour contributes a factor proportional to $\langle
n_{ij}^xn_{ij}^yn_{ij}^xn_{ij}^y\rangle_{EX}=0.06944$. This value should be
affected by the disorder and thus should vary with $\sigma$. However, it is
evident that this value can vary only in the very narrow range from $0.06944$
(at $\sigma=0$) to $0.06667$ (at $\sigma=\infty$). Therefore, the variation of
this term with $\sigma$ does not lead to a significant change in the shear
modulus as a function of $\sigma$, also because the $\sigma$ dependence of the
"isostatic" terms is comparatively much stronger. This consideration justifies
our setting $\langle
n_{ij}^xn_{ij}^yn_{ij}^xn_{ij}^y\rangle_{EX}=const=0.06944$ independent of
$\sigma$ in the following calculations.

With this decomposition, we can proportionally split the affine part of the
shear modulus as ${G^A} = G_{iso}^A + G_{ex}^A$ where $G_{iso}^A \equiv
\frac{{_1}}{{^2}}(N/V)\kappa R_0^2[6/z]z{\langle
n_{ij}^xn_{ij}^yn_{ij}^xn_{ij}^y\rangle _{GD}}$ is due to the isostatic bonds
which occur with frequency $6/z$, and
$G_{ex}^A \equiv \frac{{_1}}{{^2}}(N/V)\kappa R_0^2[(z - 6)/z]z{\langle
n_{ij}^xn_{ij}^yn_{ij}^xn_{ij}^y\rangle _{UD}}$  is due to the excess bonds that
occur with frequency $(z-6)/z$. We can similarly split the non-affine
contribution into two parts, one due to the $6$ isostatic bonds and the second
due to the additional ($z-6$) randomly placed TBN bonds:
\begin{eqnarray}
&&G^{NA} = G_{iso}^{NA} + G_{ex}^{NA} = \frac{{1}}{{2}}(N/V)\kappa R_0^26
\label{eq:Gnnn}  \\
&&\times\left\{ {{\frac{6}{z}}{{\left\langle {n_{ij}^xn_{ij}^yn_{ij}^xn_{ij}^y}
\right\rangle }_{GD}} + 3{\frac{{z - 6}}{z}}{{\left\langle
{n_{ij}^xn_{ij}^yn_{ij}^xn_{ij}^y} \right\rangle }_{EX}}} \right\}
\nonumber
\end{eqnarray}

We also need to consider a surface term to account for the fact that a
finite-size system has the atoms on the surface being displaced affinely as
they are constrained to follow the deformation imposed on the box. The fraction
of atoms on the surface scales with the total number of atoms as
$~{N^{ - 2/3}}$ (i.e. sub-extensively). It is difficult to precisely determine
the number of atoms (the thickness of the surface layer) which follow the
strain of the box in a purely affine way. To account for this uncertainty, we
introduce a parameter $B$, of order of unity, and write this additional affine
surface term as %
$G_S^A = \frac{{_1}}{{^2}}(N/V)\kappa R_0^2 zB{N^{ - 2/3}}{\langle
n_{ij}^xn_{ij}^yn_{ij}^xn_{ij}^y\rangle _{GD}}$.
With all these additional contributions put together, the total shear modulus
takes the form:
\begin{widetext}
\begin{equation}
\begin{split}
G &= \left\{ {G_{iso}^A + G_{ex}^A - G_{iso}^{NA} - G_{ex}^{NA}} \right\} +
G_S^A\\
 &= \frac{{1}}{{2}}(N/V)\kappa R_0^2\left\{ {\left[ {\frac{6}{z}} \right](z -
6){{\langle n_{ij}^xn_{ij}^yn_{ij}^xn_{ij}^y\rangle }_{GD}} + 3\frac{{{{(z -
6)}^2}}}{z}{{\langle n_{ij}^xn_{ij}^yn_{ij}^xn_{ij}^y\rangle }_{EX}}} \right\}
\\
 &+ \frac{{1}}{{2}}(N/V)\kappa R_0^2zB{N^{ - 2/3}}{\langle
n_{ij}^xn_{ij}^yn_{ij}^xn_{ij}^y\rangle _{GD}} .
\end{split}
\label{eq:Gfull}
\end{equation}
\end{widetext}
When the number of atoms N is not infinite, the most interesting situation
arises for marginal solids as $z\rightarrow6$. In this limit, and in the limit
of a large solid ($N \gg 1$) where we can assume that the surface term is
negligible, we can also neglect the quadratic term $(z-6)^2$ and have,
approximately, $G \approx (N/V)\kappa R_0^2(z - 6){\langle
n_{ij}^xn_{ij}^yn_{ij}^xn_{ij}^y\rangle _{GD}}$ for the non-affine marginal
solid with a finite degree of disorder measured by $\sigma$. Guided by the
Fig.2, we can split the disorder spectrum into a weak disorder regime with
$0<\sigma<0.1$, and a strong disorder regime with  $\sigma > 1$. In the strong
disorder regime the shear modulus is practically constant. Remarkably, in this
limit our theory recovers the result obtained independently from the numerical
analysis of random packings~\cite{ohern} and depleted regular
lattices~\cite{thorpePRB}: $G\approx (1/30)(N/V)\kappa R_0^2(z - 6)$.

On the other hand, for intermediate and weak disorder the shear modulus is
strongly dependent upon the degree of disorder. For weakly disordered cubic
networks, with $0<\sigma<0.1$, the shear modulus is proportional to $\sigma^2$:
\begin{equation}
G \approx 0.175\frac{N}{V}\kappa R_0^2(z - 6){\sigma ^2}
\label{5}
\end{equation}
If the underlying isostatic lattice is perfectly ordered, i.e.  $\sigma=0$, the
general Eq. (\ref{eq:Gfull}) has the only contribution arising from the excess
bonds, giving $G = {G_{ex}} \propto {(z - 6)^2}$. Remarkably, this scaling
agrees with the recent result obtained from the numerical study of square
lattices with randomly placed extra springs~\cite{lubensky_2010}. We have now
seen that the theory predicts that the rigidity of partially disordered lattices
near the isostatic point increases upon increasing the degree of structural
disorder.
\begin{figure}[h]
\includegraphics[width=0.95\linewidth]{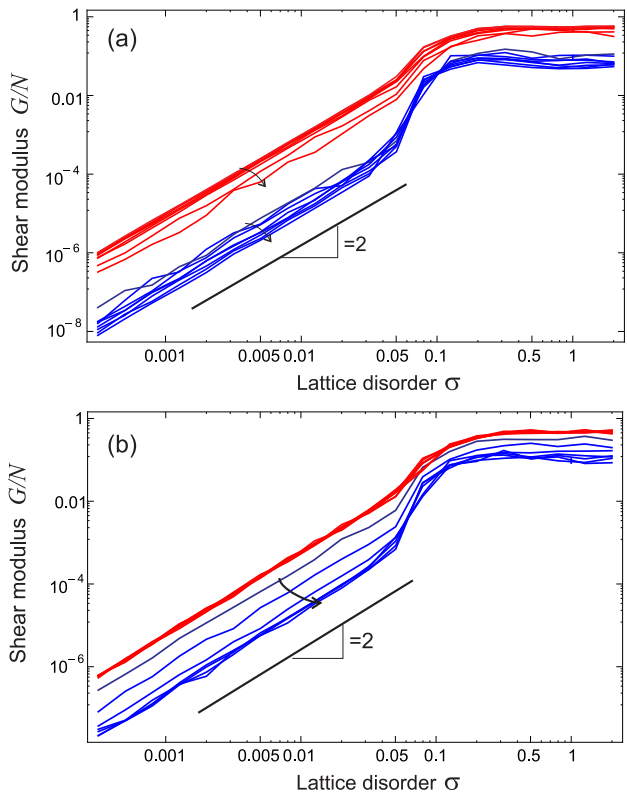}
\caption{The shear modulus of simulated isostatic networks ($z = 6$) as a
function of the disorder variance $\sigma$. Red lines: affine, blue lines:
non-affine results. In plot (a) the arrows show the curves changing with the
increasing network size, $N=4^3, 5^3, 6^3, 7^3, 8^3, 9^3$ and $10^3$,
respectively. In plot (b) the arrows show the convergence of simulation (N=216
as an example) for increasing number of steps, $t=10, 30, 100, 1000, 10^4,
3·10^4$ and $10^5$.}
\label{fig3}
\end{figure}

\section{Numerical simulations }
To test this conclusion, we performed numerical simulations where the disorder
is introduced as a perturbation on a reference cubic lattice (Fig.1c). The
magnitude of random perturbation for each atom follows the Gaussian
distribution with a variance $\sigma^2$ along each of the three Cartesian axes.
The parameter $\sigma$, therefore, controls the degree of quenched disorder in
the network, with  $\sigma=0$ corresponding to the cubic lattice and
$\sigma\geq 1$  corresponding to a completely disordered lattice. The network
topology is formed by introducing $z$ bonds placed at random on the nearest
available atoms iteratively, until each atom has $z$ neighbors. A potential
energy quadratic in the displacement is associated with every bond and the
total free energy is computed by adding the contributions from all the bonds.
{One should note that the perturbation of the lattice leads to bond lengths
that differ from the equilibrium length $R_{0}$, an effect which is not
considered by the theory. However, the contributions from all the bonds are
summed up and therefore the contributions of shorter and longer (than $R_{0}$)
bonds compensate each other yielding something not dissimilar from the
equilibrium bond length upon summing up the contributions. A further
consequence of assuming that all bonds are at the equilibrium length is the
automatic vanishing of the first-order terms in the free-energy Born expansion.
These terms are related to internal stresses but also in this case, upon
averaging over the whole network, the various contributions cancel to give a
small net contribution (a well-documented fact in the glass simulation
literature, see e.g. Tanguy et al.~\cite{barrat}). In weekly connected systems
(where $z<6$ and bond-bending interactions are active) the situation might be
different as discussed in earlier works (e.g.~\cite{alexander}).}

Upon submitting the system to small external deformations of simple shear, the
mechanical equilibrium condition on each atom/node is enforced such that it
follows the imposed deformation by moving along paths of minimum mechanical
energy. This implies, in turn, that each atom moves non-affinely in response to
the external strain (see Appendix C for detail). From the total free energy
computed in this way we measure the shear modulus of the lattices upon varying
the degree of disorder $\sigma$ and the connectivity $z$. {The results are
reported in Fig.3. First of all, these plots confirm that our simulations are
valid, both in terms of the scaling $G/N\sim N^{-2/3}$ of Eq.(\ref{eq:Gfull})
(Fig.3a) and the convergence to equilibrium (Fig.3b) tested. }It is especially
instructive to note that when only a short time is allowed for force
relaxation, the non-affine modulus is very close to the affine result.

\begin{figure}[h]
\includegraphics[width=0.95\linewidth]{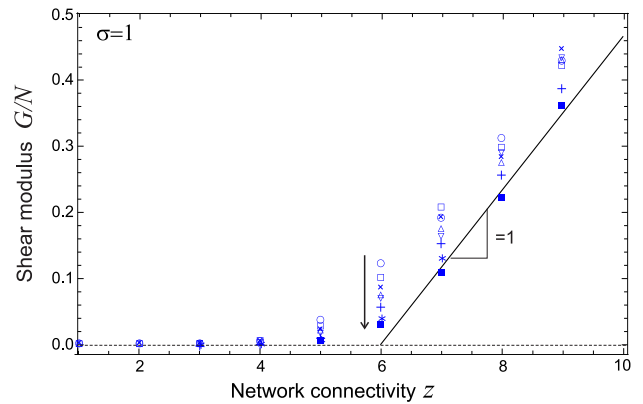}
\caption{The shear modulus of simulated networks, for a fixed value of the
disorder variance $\sigma=1$, plotted as a function of connectivity $z$, taking
only integer values in this simulation. The arrow indicates the direction of
increasing number of particles: the lattice size from top to bottom is $N=4^3,
5^3, 6^3, 8^3, 10^3, 12^3, 15^3$ and $20^3$. The solid line is the theoretical
scaling~\cite{zaccone} $G\sim (z-6)$, valid for $z\rightarrow6$.}
\label{fig4}
\end{figure}

The simulation results reproduce the fundamental laws predicted by the theory,
as summarized in Eq. (\ref{eq:Gfull}). For the isostatic (marginal-solid) limit
with $z=6$, and for low to moderate disorder, the shear modulus increases in
proportion to $\sigma^2$, and then reaches a strong-disorder plateau at
$\sigma\approx 1$. Further, it is seen that non-affinity causes a substantial
softening of the solid (lower shear modulus) with respect to the purely affine
case even though the dependence upon $\sigma$ is the same for both affine and
non-affine simulations.
Varying the connectivity at a fixed degree of disorder, the simulated networks
display a rigidity transition at the isostatic point. As shown in Fig.4, the
fundamental scaling $G \propto (z - 6)$ due to non-affinity is asymptotically
approached in the large $N$ limit. In Fig.5 theory and simulation results are
shown together as a function of the degree of disorder upon varying the
connectivity. The theory is in fairly good agreement with the simulations for
both $z=6$ and $z=7$. It should be noted that the surface term ($G_S^A$ in Eq.
(\ref{eq:Gfull})) plays an essential role in this comparison because it ensures
that the solid remains \emph{marginally} rigid with $N$ finite also at $z=6$.
Further, it is seen that upon increasing the connectivity $z$ above the
marginal-solid limit ($z=6$) the shear modulus becomes gradually less sensitive to
disorder and eventually becomes independent of $\sigma$. This is another
important and new result: the rigidity of central-force lattices with $z \geq
7$ is very little affected by the degree of structural order/disorder of the
lattice.
\begin{figure}[h]
\includegraphics[width=0.95\linewidth]{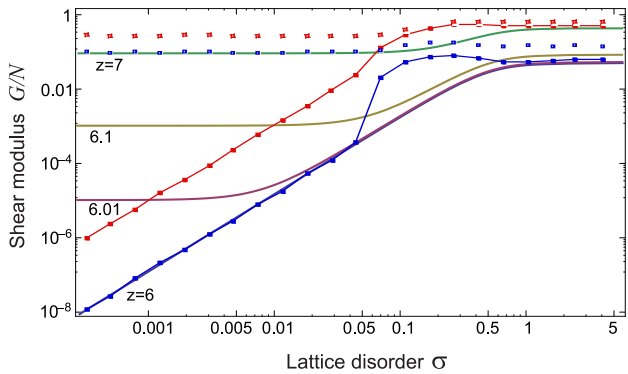}
\caption{The shear modulus from the analytical theory (solid lines) and
simulations (open symbols) upon varying both disorder variance, $\sigma$, and
lattice connectivity, $z$. Red symbols refer to the affine moduli, blue symbols
to the moduli accounting for non-affinity, for the two discrete cases $z=6$ and
$z=7$. The continuous variation of $z$ above the isostatic point $z=6$ shows
how the effect of disorder diminishes, and gradually disappears at higher
connectivity.}
\label{fig5}
\end{figure}
{In the window of disorder $0.05<\sigma<0.5$ the theory appears to
underestimate the simulation data which exhibit a large hump before reaching
the plateau corresponding to strong disorder. Although there can be many
reasons for this disagreement, the simulation being completely independent, and
using a subtly different way to define disorder $\sigma$, one possibility also
lies in the assumption (Appendix A) that the submatrices of the Hessian which
determine the form of the eigenvectors (required to calculate the non-affine
term) reduce to identity matrices upon averaging, which is an exact procedure
only for the two opposite SC and UD limits.}
We have thus shown that the rigidity of a cubic lattice increases upon
introducing structural bond-orientational disorder, with a law of direct
proportionality to the disorder variance. Clearly, the disorder-induced
stiffening is an effect controlled by the orientation average
$\langle n_{ij}^xn_{ij}^yn_{ij}^xn_{ij}^y\rangle$, which has a simple
geometrical meaning: it measures the extent to which the bond-vectors of the
lattice are misaligned with respect to the shear-deformation axes, $x$ and $y$.
In fact, the product $\langle n_{ij}^xn_{ij}^yn_{ij}^xn_{ij}^y\rangle$ vanishes
for the simple cubic lattice where all bonds are parallel to either the $x$ or
the $y$ axis. On the other hand, the product $\langle
n_{ij}^xn_{ij}^yn_{ij}^xn_{ij}^y\rangle$ is maximized and saturates to its
highest value (=1/15) when the bonds tend to be randomly oriented. In the cubic
lattice the vanishing of $\langle n_{ij}^xn_{ij}^yn_{ij}^xn_{ij}^y\rangle$ and
the shear modulus with it, reflects the fact that there is no restoring force
if all parallel columns in either the x or y direction are tilted (maintaining
the constant separation of neighbors), which is an example of a floppy mode. In
disordered lattices, with bonds misaligned with respect to x and y, the
longitudinal components of the bond displacements cannot be neglected (they are
no longer second order with respect to the displacement magnitude) which
implies a finite restoring force and thus a finite shear modulus. From a
different perspective, the factor $\langle
n_{ij}^xn_{ij}^yn_{ij}^xn_{ij}^y\rangle$ measures the average local (geometric)
frustration of the bonds and it takes a finite value whenever the bonds no
longer can be displaced without paying a finite energy cost.

\section{Conclusions}
First of all, the conclusion that bond-orientational disorder causes stiffening
of an ordered lattice can be extended to other crystal lattices by
re-evaluating the factor $\langle n_{ij}^xn_{ij}^yn_{ij}^xn_{ij}^y\rangle$. For
example, for the hexagonal close packed (hcp) lattice with $z_{HCP} =12$ we
find that
\begin{equation}
{\langle n_{ij}^xn_{ij}^yn_{ij}^xn_{ij}^y\rangle _{HCP}} = \frac{1}{{16\pi
}}\left( {\frac{{27\sqrt 3  + 3}}{{512}}} \right) \approx 0.0019 .
 \label{eq:hcp}
 \end{equation}
As a result, the ratio between the hcp modulus and the modulus of a uniformly
disordered z-coordinated lattice takes the form:
\begin{equation}
\frac{G_{HCP}}{G_{UD}} = \frac{\langle n_{ij}^xn_{ij}^yn_{ij}^xn_{ij}^y\rangle
_{HCP}}{\langle n_{ij}^xn_{ij}^yn_{ij}^xn_{ij}^y\rangle _{UD}} \approx
0.0285\frac{z_{HCP}}{z - 6},
\label{eq:Ghcp}
\end{equation}
where we used that ${G_{NA}} \equiv 0$ for the ideal hcp lattice. If, for
instance, the disordered lattice has $z=7$, then ${G_{HCP}}/{G_{UD}} \approx
1/3$, i.e. the shear modulus of an ideal hcp lattice with $z=12$ (for which the
non-affinity is certainly negligible) is almost three times smaller than the
shear modulus of a lattice with uniform disorder and $z=7$, in spite of the hcp
lattice being more strongly connected. The difference becomes more pronounced
upon increasing the coordination of the disordered lattice. Therefore the
disorder-induced stiffening of crystal lattices is a general effect not limited
to the cubic lattice studied in detail above.

Secondly, these findings may have consequences for the glass transition of
supercooled liquids. Descriptive theories of the glass transition have
emphasized the role of local icosahedral symmetry which emerges at the glass
transition~\cite{nelson}. Icosahedral clusters have also been identified with
the cooperatively rearranging regions of mean-field (e.g. Adam-Gibbs)
theories~\cite{debenedetti_book}. The growth and jamming of such icosahedral
clusters lead to the sudden increase of the viscosity and eventually to the
structural arrest. In the icosahedral clusters an atom at the center is bonded
to its $12$ nearest neighbors, forming an icosahedron where none of the bonds
can be parallel with either the $x$ or $y$ directions. Instead, clusters with
hcp symmetry could be formed which have the same number of bonds as the
icosahedra but 6 of these bonds can be aligned with the $x$ and $y$ axis. The
same applies to fcc clusters. Therefore it follows that the factor $\langle
n_{ij}^xn_{ij}^yn_{ij}^xn_{ij}^y\rangle$ is much larger for the icosahedra than
for either hcp or fcc clusters and thus the icosahedra are more rigid than the
hcp or fcc clusters. This implies that, during the supercooling process, the
icosahedra, which are the precursor of the rigid glass state~\cite{andersen},
are able to stand the internal stresses generated by the
vitrification~\cite{alexander} better than the hcp or fcc clusters which are
the precursor of the ordered crystal. This is an unprecedented observation that
sheds light on the physical origin of the onset of rigidity at the glass
transition.

In further developments, this theory can be applied to analyze the onset of
rigidity in systems which possess medium-range order~\cite{angell}, such as the
technologically important amorphous semiconductors and metallic
glasses~\cite{phillips,sheng}. {In principle, the orientation distribution
function for partially disordered lattices (e.g., those containing defects such
as dislocations causing distortion of the bond-orientations, or the above
mentioned solids with medium-range order) can be extracted from structural
investigation of the material in question (e.g., scattering or microscopy
techniques) and used as input in our theory to investigate the elastic
properties as a function of structure. }

In summary, by combining theory and simulations we have established a
fundamental law which governs the rigidity of solids as a function of
structural disorder. According to this law, the stiffness of lattices increases
with the degree of bond-orientational disorder. Our finding suggests a new way
of increasing the mechanical stability of technologically important
marginal-solids such as e.g. Po$^{210}$ and the superconducting Ca-III, both of
which exhibit simple cubic lattices which therefore can be stabilized by
introducing structural disorder (e.g. in the form of defects). Furthermore, our
findings suggest a new way of looking at the unresolved problem of the glass
transition. In light of our results, vitrifying liquids upon crossing the
marginal-solidity or isostatic threshold find themselves in disordered
configurations which are intrinsically more rigid than the ordered ones.

\subsection*{Acknowledgments.}
This research has been supported by the Swiss National Foundation
(PBEZP2-131153) and the EPSRC TCM Programme Grant. We appreciate useful
discussions and other input from S.D. Guest, T.C. Lubensky, M. Warner, H. J.
Herrmann, and R. Blumenfeld.\\

\appendix
\section{Non-affine deformations}
In the following we give a brief outline of the analytical method that we
devised to calculate the elastic constants of disordered harmonic lattices
which accounts for non-affinity\cite{lemaitre,zaccone}. The method makes use of
the non-affine linear response formalism for disordered solids which is
described in detail by Lemaitre and Maloney\cite{lemaitre}. One of the central
results of the formalism, which represents the starting point of our
derivation, is the exact lattice sum for the elastic moduli accounting for
non-affinity\cite{lemaitre}
\begin{eqnarray} \label{S1}
&&{C_{\iota \xi \kappa \chi }} = C_{\iota \xi \kappa \chi }^A - C_{\iota \xi
\kappa \chi }^{NA} \\
&=& \frac{{R_0^2\kappa }}{{2V}}\sum\limits_{ij} {{c_{ij}}} n_{ij}^\iota
n_{ij}^\kappa n_{ij}^\xi n_{ij}^\chi  - \frac{1}{V}\sum\limits_k^{3N}
{\frac{{({\underline{\Xi}_{\iota \xi }} \cdot
{\underline{\mathbf{v}}_k}{\rm{)}}({\underline{\Xi}_{\kappa \chi }} \cdot
{\underline{\mathbf{v}}_k}{\rm{)}}}}{{{\lambda _k}}}} \nonumber
\end{eqnarray}
The superscripts A and NA denote the affine and non-affine parts of the moduli,
respectively. The affine part is the standard Born-Huang expression in terms of
a lattice sum over nearest-neighbors (NN) $i$ and $j$, where
${\underline{n}_{ij}} = (\cos \phi \sin \theta ,\sin \phi \sin \theta ,\cos
\theta )$ is the unit vector defining the orientation of bond $ij$. $c_{ij}$ is
the adjacency matrix ($c_{ij} =1$ for two NN atoms and $c_{ij}=0$ otherwise)
and $\kappa$ the harmonic spring constant of the harmonic interaction between
two NN atoms. The non-affine part is a sum over the $3N$ eigenmodes of the
Hessian or dynamical matrix of the lattice. The $3N\times3N$ Hessian matrix for
harmonic lattices is given by
\begin{equation}\label{S2}
H_{ij}^{\alpha \beta } = {\delta _{ij}}\sum\limits_s {\kappa {c_{is}}}
n_{is}^\alpha n_{is}^\beta  - (1 - {\delta _{ij}})\kappa {c_{ij}}n_{ij}^\alpha
n_{ij}^\beta
\end{equation}
Eq.(\ref{S2}) follows from replacing the harmonic potential $U({r_{ij}}) =
\frac{1}{2}\kappa {({r_{ij}} - {R_0})^2}$  in the definition of the Hessian
matrix:
$H_{ij}^{\alpha \beta } \equiv {\partial ^2}U/\partial r_i^\alpha \partial
r_j^\beta $. $R_0$ is the rest length of the bonds. $\underline{\mathbf{v}}_k$
and $\lambda_{k}$ in Eq.(\ref{S1}) are eigenvectors and eigenvalues of the
Hessian, respectively. The inner product $({\underline{\Xi}_{\iota \xi }} \cdot
{\underline{\mathbf{v}}_k}{\rm{)}}$ is the projection of the affine force field
${\underline{\Xi}_{\iota \xi }} $ (i.e., the force field exerted on every atom by
the affine motions of its neighbours) on the eigenvector $\underline{\mathbf{v}}_k$. The
analytical form of the affine fields is given by~\cite{lemaitre}:
$\Xi _{i,\kappa \chi }^\alpha  =  - \sum\limits_j {{R_{ij}}\kappa {c_{ij}}}
n_{ij}^\alpha n_{ij}^\kappa n_{ij}^\chi$. Thus, the evaluation of the
non-affine term in the elastic moduli reduces to the task of evaluating the
eigenmodes of the Hessian,
$\underline{\mathbf{v}}_k=\underline{a}_{r}\bigotimes \underline{w}_l$ where
$r=1...N$ and $l=x,y,z$. In general, there are no analytical routes to evaluate
the eigenmodes. This becomes possible however if one chooses
$\underline{w}_l=\underline{e}_{l}$ where $\underline{e}_{l}$ is the standard
Cartesian basis of $\mathbb{R}^{3}$, as we show below. This treatment is exact
in the case of the simple cubic (SC) lattice where the Hessian is exactly given
by
$H_{ij}^{\alpha\beta}=\frac{\kappa}{3}\left(\delta_{ij}\sum_j c_{ij}-
(1-\delta_{ij})c_{ij}\right)\delta_{\alpha\beta} $ and clearly the
$\underline{\mathbf{v}}_k=\underline{a}_{r}\bigotimes \underline{e}_l$ are
eigenvectors of the Hessian because the submatrices of the Hessian are
diagonal.
In the case of the uniformly disordered (UD) (isotropic) lattice as defined in
the main article, it can be shown that the eigenvectors have the same form.
A general theorem states that if ${\underline{\mathbf{v}}_k}$ and $\lambda_{k}$
are respectively the eigenvectors and eigenvalues of a matrix
$\underline{\underline{A}}$, the same eigenvectors are shared by a matrix
$f(\underline{\underline{A}})$, for any function $f$ of the matrix. The
eigenvalue equation for the latter matrix is then given by:
$f(\underline{\underline{A}}){\underline{\mathbf{v}}_k}=f(\lambda_{k}){\underline{\mathbf{v}}_k}$.
Let us take $f(...)=\langle...\rangle$, where $\langle...\rangle$ denotes the
average over the bond-orientational disorder, as usual. Then we have
$f(\underline{\underline{H}})=\frac{\kappa}{3}\left(\delta_{ij}\sum_j c_{ij}-
(1-\delta_{ij})c_{ij}\right)\delta_{\alpha\beta}$, where $\delta_{\alpha\beta}$
is the Kronecker's delta. Furthermore, $\lambda_{k}=m\omega_{k}^{2}$ where
$\omega_{k}$ are the normal mode frequencies of the solid. The lattice with
uniform disorder is an isotropic solid, which means that the normal mode
frequencies depend only on scalar quantities. Clearly the same applies to the
$\lambda_{k}$ which in turn implies: $f(\lambda_{k})=\lambda_{k}$. This result
establishes that the eigenvectors of the form:
$\underline{\mathbf{v}}_k=\underline{a}_{r}\bigotimes \underline{e}_l$ are also
eigenvectors of the Hessian for the uniformly disordered lattice.
Hence, in both the SC and UD limits one can write the eigenvalue equation for
the Hessian as: $(\underline{\underline{\tilde
H}}\otimes\underline{\underline{I}})(\underline{a}\otimes\underline{e})=\lambda(\underline{a}\otimes\underline{e})$,
where $\underline{\underline{\tilde
H}}\equiv\frac{\kappa}{3}\left(\delta_{ij}\sum_j c_{ij}-
(1-\delta_{ij})c_{ij}\right)$ and $\underline{\underline{I}}$ denotes the
$3\times3$ identity matrix.
Then the inner products of the affine fields with the eigenmodes
become\cite{zaccone}:
\begin{eqnarray}\label{S3}
&&({\underline{\Xi}_{\iota \xi }} \cdot {\underline{\mathbf{v}}_k}{\rm{)}} = \\
&=& {\kappa ^2}R_0^2\sum\limits_{rs,r's'}^{}
{{a_r}{a_{r'}}{c_{rs}}{c_{r's'}}n_{rs}^l} n_{rs}^\iota n_{rs}^\xi
n_{r's'}^ln_{r's'}^\kappa n_{r's'}^\chi    \nonumber
\end{eqnarray}
where the sum runs over two pairs of NN atoms at the time, $rs$ and $r's'$.
Consistent with the main approximation of this theory, we replace the
orientation-dependent terms with their orientation average. One finds that
$\langle {n_{rs}^ln_{rs}^\iota n_{rs}^\xi n_{r's'}^ln_{r's'}^\kappa
n_{r's'}^\chi } \rangle  = ({\delta _{rr'}}{\delta _{ss'}} - {\delta
_{rs'}}{\delta _{sr'}}){B_{l,\iota \xi \kappa \chi }}$ where ${B_{l,\iota \xi
\kappa \chi }} = \langle {n_{ij}^\iota n_{ij}^\xi n_{ij}^\kappa n_{ij}^\chi }
\rangle$ and $\langle ...\rangle  = \int {...f(\theta ,\phi )} \sin \theta
d\theta d\phi$ denotes the average over bond orientations according to some
orientation distribution function $f(\theta ,\phi )$. Replacing this average in
Eq.(\ref{S3}), one obtains:
\begin{eqnarray}\label{S4}
&& {({\underline{\Xi}_{\iota \xi }} \cdot
{\underline{\mathbf{v}}_k}{\rm{)}}({\underline{\Xi}_{\kappa \chi }} \cdot
{\underline{\mathbf{v}}_k}{\rm{)}}} = \\
&=& {\kappa ^2}R_0^2{B_{l,\iota \xi \kappa \chi }}\left( {\sum\limits_{rs}
{a_r^2{c_{rs}}{c_{rs}} - \sum\limits_{rs} {a_r^{}a_s^{}{c_{rs}}{c_{sr}}} } }
\right)  \nonumber
\end{eqnarray}
Furthermore, we have that $c_{rs}^2 = {c_{rs}}{c_{sr}} = {c_{rs}}$ and the
identities
$\sum\limits_r^N {a_r^2} \sum\limits_s {{c_{rs}}}  - \sum\limits_{rs} {{a_r}}
{a_s}{c_{rs}} = \sum\limits_{rs}^N {{a_r}} {a_s}[(\sum\limits_j^N {{c_{rj}}}
){\delta _{rs}} - {c_{rs}}(1 - {\delta _{rs}})] = \frac{3}{\kappa
}\sum\limits_{rs}^N {{a_r}} {a_s}{\tilde H_{rs}}$, where we defined
$\underline{\underline{H}} = \tilde{\underline{\underline{H}}} \otimes
\underline{\underline{I}}$
and $\underline{\underline{I}}$  is the $3\times3$ identity matrix. Recalling
that $\sum\limits_s^N {{{\tilde H}_{rs}}} {\mkern 1mu} {a_s} = \lambda {a_r}$
we obtain $ {({\underline{\Xi}_{\iota \xi }} \cdot
{\underline{\mathbf{v}}_k}{\rm{)}}({\underline{\Xi}_{\kappa \chi }} \cdot
{\underline{\mathbf{v}}_k}{\rm{)}}}= \kappa R_0^2{\mkern 1mu} {\lambda
_k}{B_{l,\iota \xi \kappa \chi }}$
Using this result in Eq.(\ref{S1}), the non-affine part of the shear modulus
follows as:
\begin{equation}\label{S5}
\begin{aligned}
{G^{NA}} &\equiv \frac{1}{V}\sum\limits_k^{3N} {\frac{{({{\underline \Xi
}_{xy}} \cdot {{\underline {\rm{v}} }_k}{\rm{)}}({{\underline \Xi  }_{xy}}
\cdot {{\underline {\rm{v}} }_k}{\rm{)}}}}{{{\lambda _k}}}} \\
&= \frac{1}{V}\sum\limits_{k = 1}^N {\sum\limits_{l = 1}^3 {\mkern 1mu}  }
\frac{{3\kappa R_0^2{\mkern 1mu} {\lambda _k}{\mkern 1mu} {B_{l,\iota \xi
\kappa \chi }}}}{{{\lambda _k}}}\\
&= 3N\kappa R_0^2{\mkern 1mu} \langle n_{ij}^xn_{ij}^yn_{ij}^xn_{ij}^y\rangle
/V
\end{aligned}
\end{equation}
which is a key result used in the main part of the article.

{This theory strictly applies to the SC and UD limits only because only in
these specific cases the angular average $\langle n_{ij}^\alpha
n_{ij}^\beta\rangle=\delta_{\alpha\beta}/3$ is exact. However, we make the
additional assumption that for the intermediate degrees of disorder this result
still holds approximately, i.e. $\langle n_{ij}^\alpha
n_{ij}^\beta\rangle\simeq\delta_{\alpha\beta}/3$ for $0<\sigma<\infty$. Within
this approximation Eq.(\ref{S5}) can be used to evaluate the elastic moduli of
lattices with disorder (parameterized by variable $\sigma$), by using the
Gaussian ODF defined in the main article to evaluate the orientation average in
Eq.(\ref{S5}). In the worst situation, this approximation still should provide
a reasonable interpolation since it correctly describes both the lower (SC) and
upper (UD) limits of the disorder spectrum. The validity of the approximation
has been checked by comparison with simulation data in the main article and as
expected it provides a good description of the data apart for a window of
disorder $0.05<\sigma<0.5$ located in the middle of the spectrum.}

\section{Gaussian disorder}
The distribution function for the orientation of the lattice bonds (ODF)
$f(\theta ,\phi )$ is defined by the following relation
\begin{equation}\label{S6}
\langle ...\rangle  = \int {...f(\theta ,\phi )} \sin \theta d\theta d\phi
\end{equation}
To evaluate the shear modulus we need to evaluate its structure-dependent part
\begin{equation}\label{S7}
\langle n_{ij}^xn_{ij}^yn_{ij}^xn_{ij}^y\rangle  = \int {f(\theta ,\phi )}
{\sin ^4}\theta {\cos ^2}\phi {\sin ^2}\phi \sin \theta d\theta d\phi
\end{equation}
All the information about the structure of the lattice is contained in
$f(\theta ,\phi )$. For isostatic lattices in $d=3$, as explained in the main
body of the article, the degree of structural disorder can be varied
continuously from the simple cubic lattice to the complete uncorrelated
disorder by using the following Gaussian ODF
\begin{equation}\label{S8}
{f_{GD}}(\theta ,\phi ) = \frac{2}{6}f_{GD}^{ze}(\theta ,\phi ) +
\frac{4}{6}f_{GD}^{az}(\theta ,\phi )
\end{equation}
with
\begin{equation}\label{S9}
\begin{array}{l}
f_{GD}^{ze}(\theta ,\phi ) = {N_{ze}}({\sigma ^2}){e^{ - {\theta ^2}/2{\sigma
^2}}}\\
f_{GD}^{az}(\theta ,\phi ) = {N_{az}}({\sigma ^2}){e^{ - {{(\theta  - \pi
/2)}^2}/2{\sigma ^2}}}{e^{ - {\phi ^2}/2{\sigma ^2}}}
\end{array}
\end{equation}
The normalization factors ${N_{ze}}$ and ${N_{az}}$ are defined by
\begin{equation}\label{S10}
\begin{array}{l}
\int {{N_{ze}}({\sigma ^2})f_{GD}^{ze}(\theta ,\phi )\sin \theta d\theta d\phi
= 1} \\
\int {{N_{az}}({\sigma ^2})f_{GD}^{az}(\theta ,\phi )\sin \theta d\theta d\phi
= 1}
\end{array}
\end{equation}
Combining Eq.(\ref{S8})-(\ref{S10}) we obtain the following analytical
expression for the microstructure-dependent factor in the shear modulus:
\begin{widetext}
\begin{equation}\label{S11}
\begin{array}{l}
{\langle n_{ij}^xn_{ij}^yn_{ij}^xn_{ij}^y\rangle _{GD}} = \int {{f_{GD}}(\theta
,\phi )} {\sin ^4}\theta {\cos ^2}\phi {\sin ^2}\phi \sin \theta d\theta d\phi
\\
{\rm{                       = \{ }}{{\rm{e}}^{ - {\rm{12}}{\sigma ^2}}}\sqrt
{{\rm{1/}}{\sigma ^2}} \sqrt {{\sigma ^2}} {\rm{ (Erf [(}}\pi  - {\rm{5i
}}{\sigma ^2}{\rm{)}}/\sqrt {{\rm{2}}{\sigma ^2}} {\rm{]  +  Erf [(}}\pi {\rm{
+  5i}}{\sigma ^2}{\rm{)/}}\sqrt {{\rm{2}}{\sigma ^2}} {\rm{]  }}\\
{\rm{                          + 5}}{{\rm{e}}^{{\rm{8}}{\sigma ^2}}}{\rm{(Erf
[(}}\pi  - {\rm{3i}}{\sigma ^2}{\rm{)/}}\sqrt {{\rm{2}}{\sigma ^2}} {\rm{] }} -
{\rm{Erf [(}}\pi  + {\rm{3i}}{\sigma ^2}{\rm{)/}}\sqrt {{\rm{2}}{\sigma ^2}}
{\rm{]}}\\
{\rm{                        }} - {\rm{2 }}{{\rm{e}}^{{\rm{4}}{\sigma
^2}}}{\rm{(Erf [(}}\pi  - {\rm{i}}{\sigma ^2}{\rm{)/}}\sqrt {{\rm{2}}{\sigma
^2}} {\rm{]}} - {\rm{Erf[(}}\pi {\rm{  +  i}}{\sigma ^2}{\rm{)/}}\sqrt
{{\rm{2}}{\sigma ^2}} {\rm{]  +  2iErfi[}}\sqrt {{\sigma ^2}{\rm{/2}}} {\rm{])
}}\\
{\rm{                          + 2i Erfi[(3 }}\sqrt {{\sigma ^2}}
{\rm{)/}}\sqrt 2 {\rm{])}} - {\rm{2 i Erfi[5 }}\sqrt {{\sigma ^2}}
{\rm{/}}\sqrt 2 {\rm{])\} }}\\
{\rm{                         }} \times {\rm{\{ 384(Erf[((}}\pi {\rm{  +
i}}{\sigma ^2}{\rm{) }}\sqrt {1/2{\sigma ^2}} {\rm{]}} - {\rm{2 iErfi[1/}}\sqrt
{2/{\sigma ^2}} {\rm{]  }}\\
{\rm{                          + iErfi[(}}\sqrt {1/{\sigma ^2}} {\rm{(i }}\pi
{\rm{  +  }}{\sigma ^2}{\rm{))/}}\sqrt 2 {\rm{])}}{{\rm{\} }}^{ - 1}}\\
{\rm{                        }} + {\rm{\{ }}{{\rm{e}}^{{\rm{ - 20}}{\sigma
^2}}}{\rm{(5}}{{\rm{e}}^{{\rm{8}}{\sigma
^2}}}{\rm{(2}}{{\rm{e}}^{{\rm{4}}{\sigma ^2}}}{\rm{(Erf[(}}\pi  - {\rm{2
i}}{\sigma ^2}{\rm{)/2}}\sqrt {2{\sigma ^2}} {\rm{]  +  Erf[(}}\pi {\rm{ +
2i}}{\sigma ^2}{\rm{)/2}}\sqrt {2{\sigma ^2}} {\rm{]) }}\\
{\rm{                          +  Erf[(}}\pi  - {\rm{6i}}{\sigma ^2}{\rm{)/(2
}}\sqrt {2{\sigma ^2}} {\rm{)]  +  Erf[(}}\pi {\rm{ +  6 i}}{\sigma
^2}{\rm{)/(2}}\sqrt {2{\sigma ^2}} {\rm{)]) }}\\
{\rm{                          + Erf[(}}\pi  - {\rm{10 i}}{\sigma
^2}{\rm{)/(2}}\sqrt {2{\sigma ^2}} {\rm{)]  }}\\
{\rm{                          + Erf[(}}\pi {\rm{ +  10 i}}{\sigma ^2}{\rm{)/(2
}}\sqrt {2{\sigma ^2}} {\rm{)])(}} - {\rm{2  +  2 }}{{\rm{e}}^{{\rm{8}}{\sigma
^2}}}{\rm{Erf[(}}\sqrt 2 \pi {\rm{)/}}\sqrt {{\sigma ^2}} {\rm{]  }}\\
{\rm{                          + Erfc[(}}\sqrt 2 {\rm{(}}\pi  - {\rm{2
i}}{\sigma ^2}{\rm{))/}}\sqrt {{\sigma ^2}} {\rm{]  +  Erfc[(}}\sqrt 2
{\rm{(}}\pi {\rm{  +  2 i}}{\sigma ^2}{\rm{))/}}\sqrt {{\sigma ^2}} {\rm{]))\}
}}\\
{\rm{                        }} \times {\rm{\{ 768 Erf[}}\pi {\rm{/}}\sqrt
{2{\sigma ^2}} {\rm{] (Erf[(}}\pi  - {\rm{2 i}}{\sigma ^2}{\rm{)/2}}\sqrt
{2{\sigma ^2}} {\rm{]  }}\\
{\rm{                         + Erf[(}}\pi {\rm{  +  2 i}}{\sigma ^2}{\rm{)/(2
}}\sqrt {2{\sigma ^2}} {\rm{)])}}{{\rm{\} }}^{ - 1}}
\end{array}
\end{equation}
\end{widetext}
Eq.(\ref{S11}) is a function of the Gaussian variance $\sigma^2$  only and has
been plotted in Fig.2 in the main body of the article.

\subsection*{Derivation of the Gaussian ODF}
Here we provide a proof of the expression given by Eq.(2) in the main article
(Eqs.(\ref{S8})-(\ref{S9}) in the section above) for the Gaussian ODF used to
vary the degree of disorder in our model. We start by considering the
orientations of the bonds in the primitive cell of the simple cubic (SC)
lattice. There are $4$ bonds lying in the $xy$ plane that are defined by pairs
($\theta,\phi$) of angles $(\pi/2,0)$, $(\pi/2,\pi/2)$, $(\pi/2,\pi)$, and
$(\pi/2,3\pi/2)$. Further, there are two bonds lying along the polar z-axis
defined by $(0,\forall\phi)$ and $(\pi,\forall\phi)$. The Gaussian model
realizes the situation where each of these bonds is distributed around the
simple-cubic angle pairs mentioned above according to a Gaussian distribution
$f_{GD}$ with variance $\sigma^{2}$. In the limit $\sigma\rightarrow0$ the
model develops Dirac deltas at the SC bond orientations thus recovering the
perfectly ordered SC lattice. We wish to demonstrate that in order to evaluate
the angular average $\langle n_{ij}^xn_{ij}^yn_{ij}^xn_{ij}^y\rangle_{GD}=\int
{...f_{GD}(\theta ,\phi )} \sin \theta d\theta d\phi$, one needs only to
consider the two principal orientations $(\pi/2,0)$ and $(0,\forall\phi)$ (from
which the other ones can be obtained upon application of the symmetry
operations of the group $O_h$), which we denote as \emph{azimuthal} and
\emph{zenithal} respectively, because all the other orientations contribute
terms which are identical to either of these two. In other words, we want to
show that for the purpose of evaluating $\langle
n_{ij}^xn_{ij}^yn_{ij}^xn_{ij}^y\rangle_{GD}$ the following identities hold:
\begin{equation}\label{S12}
f_{GD}\equiv
\frac{1}{6}\left(\sum_{p=0}^{3}f_{GD}^{\{\frac{\pi}{2},p{\frac{\pi}{2}}\}}+\sum_{q=0}^{1}f_{GD}^{\{q\pi,\forall\phi\}}\right)=\frac{4}{6}f_{GD}^{az}+\frac{2}{6}f_{GD}^{ze}
\end{equation}
where we identified:
\begin{equation}\label{S13}
\begin{aligned}
f_{GD}^{az}&\equiv f_{GD}^{\{\frac{\pi}{2},0\}}\equiv{N_{az}}({\sigma ^2}){e^{
- {{(\theta  - \frac{\pi }{2})}^2}/2{\sigma ^2}}}{e^{ - {\phi ^2}/2{\sigma
^2}}}\\
f_{GD}^{ze}&\equiv f_{GD}^{\{0,\forall\phi\}}\equiv{N_{ze}}({\sigma ^2}){e^{ -
{\theta ^2}/2{\sigma ^2}}}
\end{aligned}
\end{equation}
i.e. the first terms of the two sums in Eq.(\ref{S12}), defined by $p=0$ and
$q=0$, respectively.
Let us start with the orientations in the $xy$ plane. The principal azimuthal
orientation $f_{GD}^{az}$, with $p=0$, contributes to the average $\langle
n_{ij}^xn_{ij}^yn_{ij}^xn_{ij}^y\rangle_{GD}$ the following integral:
\begin{equation}\label{S14}
\begin{aligned}
&\int{\sin ^4}\theta {\cos ^2}\phi {\sin ^2}\phi
e^{-(\theta-\frac{\pi}{2})^{2}/2\sigma^{2}}e^{-\phi^{2}/2\sigma^{2}}\sin\theta
d\theta d\phi\\
&+\int{\sin ^4}\theta {\cos ^2}\phi {\sin ^2}\phi
e^{-(\theta-\frac{\pi}{2})^{2}/2\sigma^{2}}e^{-(\phi-2\pi)^{2}/2\sigma^{2}}\sin\theta
d\theta d\phi\\
&=2\int{\sin ^4}\theta {\cos ^2}\phi {\sin ^2}\phi
e^{-(\theta-\frac{\pi}{2})^{2}/2\sigma^{2}}e^{-\phi^{2}/2\sigma^{2}}\sin\theta
d\theta d\phi
\end{aligned}
\end{equation}
where we omitted the normalization factor and where the equality holds because
the functions in the two integral on the l.h.s. subtend the same area within
the same interval of integration $[0,2\pi]$. It should be noted that the second
term on the l.h.s. is strictly required in order to have a Gaussian centered on
$\phi=0$. Without this term, since integration goes from $0$ to $2\pi$, one
would have only half of the Gaussian and would not properly count the
contributions from the angles which lie close to $\phi=2\pi$ (or equivalently
on the negative axis close to $\phi=0^{-}$).

The second term in the first sum in Eq.(\ref{S12}), with $p=1$, contributes to
the average the following term:
\begin{equation}\label{S15}
\int_{0}^{\pi}{\sin ^5}\theta
e^{-(\theta-\frac{\pi}{2})^{2}/2\sigma^{2}}d\theta\int_{0}^{2\pi} {\cos ^2}\phi
{\sin ^2}\phi e^{-(\phi-\frac{\pi}{2})^{2}/2\sigma^{2}} d\phi
\end{equation}
In this integral, let us change to the variable $\Phi\equiv
\phi-\frac{\pi}{2}$. Upon applying the well-known trigonometric relations:
$\sin(\phi+\frac{\pi}{2})=\cos\phi$ and $\cos(\phi+\frac{\pi}{2})=-\sin\phi$,
the integral in Eq.(\ref{S15}) becomes:
\begin{equation}\label{S16}
\int_{0}^{\pi}{\sin ^5}\theta
e^{-(\theta-\frac{\pi}{2})^{2}/2\sigma^{2}}d\theta\int_{-\frac{\pi}{2}}^{\frac{3\pi}{2}}
{\cos ^2}\Phi {\sin ^2}\Phi e^{-\Phi^{2}/2\sigma^{2}} d\Phi
\end{equation}
This integral is identical to the r.h.s. of Eq.(\ref{S14}) because the area
subtended by the function ${\cos ^2}\phi {\sin ^2}\phi
e^{-\phi^{2}/2\sigma^{2}}$ (and obviously by ${\cos ^2}\Phi {\sin ^2}\Phi
e^{-\Phi^{2}/2\sigma^{2}}$) in the interval $[0,2\pi]$ is the same as in the
interval $[-\pi/2,3\pi/2]$, as one can easily verify by plotting the functions.
This rigorously establishes the identity $f_{GD}^{az}\equiv
f_{GD}^{\{\frac{\pi}{2},0\}}=f_{GD}^{\{\frac{\pi}{2},\frac{\pi}{2}\}}$ which
holds in the calculation of $\langle
n_{ij}^xn_{ij}^yn_{ij}^xn_{ij}^y\rangle_{GD}$. With analogous arguments it is
easy to show that $f_{GD}^{\{\frac{\pi}{2},0\}}=f_{GD}^{\{
\frac{\pi}{2},p\pi/2\}}$ also  holds $\forall p$ where $p$ is an integer. This
leads us to conclude that:
\begin{equation}\label{S17}
\sum_{p=0}^{3}f_{GD}^{\{\frac{\pi}{2},p{\frac{\pi}{2}}\}}=4f_{GD}^{az}
\end{equation}

To complete the derivation of Eq.(\ref{S12}) (i.e. Eq.(2) in the main article),
we still need to demonstrate that
$\sum_{q=0}^{1}f_{GD}^{\{q\pi,\forall\phi\}}=2f_{GD}^{ze}$ holds as well in the
calculation of the angular average. The principal zenithal orientation
$f_{GD}^{ze}$, with $q=0$, contributes to the angular average $\langle
n_{ij}^xn_{ij}^yn_{ij}^xn_{ij}^y\rangle_{GD}$ the following integral:
\begin{equation}\label{S18}
\int_{0}^{\pi}{\sin ^5}\theta e^{-\theta^{2}/2\sigma^{2}}d\theta\int_{0}^{2\pi}
{\cos ^2}\phi {\sin ^2}\phi d\phi
\end{equation}
where we omitted again the normalization factor. The orientation with $q=1$
gives the following contribution to the angular average:
\begin{equation}\label{S19}
\int_{0}^{\pi}{\sin ^5}\theta
e^{-(\theta-\pi)^{2}/2\sigma^{2}}d\theta\int_{0}^{2\pi} {\cos ^2}\phi {\sin
^2}\phi d\phi
\end{equation}
This integral, upon changing to the variable $\Theta\equiv \theta-\pi$,
becomes:
\begin{equation}\label{S20}
\int_{-\pi}^{0}-{\sin ^5}\Theta
e^{-\Theta^{2}/2\sigma^{2}}d\Theta\int_{0}^{2\pi} {\cos ^2}\phi {\sin ^2}\phi
d\phi
\end{equation}
where we used $\sin(\theta-\pi)=-\sin\theta$. The function ${\sin ^5}\Theta
e^{-\Theta^{2}/2\sigma^{2}}$ is an odd function of the variable $\Theta$. This
implies that: $\int_{-\pi}^{0}-{\sin ^5}\Theta
e^{-\Theta^{2}/2\sigma^{2}}d\Theta=\int_{0}^{\pi}{\sin ^5}\Theta
e^{-\Theta^{2}/2\sigma^{2}}d\Theta$. The last identity establishes that the
integral in Eq.(\ref{S19}) and the integral in Eq.(\ref{S18}) are identical.
This result in turn demonstrates that, for the purpose of calculating $\langle
n_{ij}^xn_{ij}^yn_{ij}^xn_{ij}^y\rangle_{GD}$:
\begin{equation}\label{21}
\sum_{q=0}^{1}f_{GD}^{\{q\pi,\forall\phi\}}=2f_{GD}^{ze}.
\end{equation}

Recollecting the results presented above, we have shown rigorously that to the
effect of calculating $\langle n_{ij}^xn_{ij}^yn_{ij}^xn_{ij}^y\rangle_{GD}$
the following equality holds:
\begin{equation}\label{22}
\begin{split}
&f_{GD}=\frac{1}{6}\left(\sum_{p=0}^{3}f_{GD}^{\{\frac{\pi}{2},p{\frac{\pi}{2}}\}}+\sum_{q=0}^{1}f_{GD}^{\{q\pi,\forall\phi\}}\right)=\frac{4}{6}f_{GD}^{az}+\frac{2}{6}f_{GD}^{ze}\\
&=\frac{4}{6}{N_{az}}({\sigma ^2}){e^{ - {{(\theta  - \frac{\pi
}{2})}^2}/2{\sigma ^2}}}{e^{ - {\phi ^2}/2{\sigma
^2}}}+\frac{2}{6}{N_{ze}}({\sigma ^2}){e^{ - {\theta ^2}/2{\sigma ^2}}}
\end{split}
\end{equation}
which has been used in the main article.

As shown in Fig.2 of the main article, using Eq.(\ref{
22}) to evaluate
$\langle n_{ij}^xn_{ij}^yn_{ij}^xn_{ij}^y\rangle_{GD}$ leads us to recover
exactly $\lim_{\sigma\rightarrow\infty}\langle
n_{ij}^xn_{ij}^yn_{ij}^xn_{ij}^y\rangle_{GD}=1/15$ which is the value that one
calculates assuming that the bond orientations are randomly distributed in the
solid angle, i.e. using the uniform-disorder ODF: $f_{UD}=1/4\pi$. This result
gives a further confirmation of the correctness of the above derivation.

\section*{Appendix C. Simulation of non-affine disordered lattices}
{We simulate the effect of non-affine deformations on the shear modulus of an
elastic solid by implementing the following scheme. First we place $N$ points
on a perfect simple cubic lattice, then we perturb the lattice topology to
induce a variable degree of disorder and we assign a harmonic elastic energy to
each bond. Finally we deform the lattice by implementing a gradient descent
method at zero-temperature to find the local energy minima. }

Typically we choose a lattice composed of $20\times20\times20$ atoms
corresponding to $N=8000$, having confirmed that the intrinsically intensive
value of the modulus has saturated on increasing $N$. Each point is then
perturbed from its lattice position by drawing three random samples from a
Gaussian distribution of variance and displacing the position of each atom by
these values in the $x, \, y$ and $z$ directions. This process is repeated
independently for each of the $N$ atoms. We enforce the condition that all
positions must be in the interval $[0,N]$ so that all positions $x,y,z$ are, in
fact, $modulo(x,N)$, $modulo(y,N)$, and $modulo(z,N)$. Once the positions of
the vertices have been defined as above, the connections in the network are
introduced by picking a vertex at random and forming bonds to each of the z
nearest neighbors, provided they do not already have z neighbors, i.e. they are
available. This process is repeated until the network is fully formed and no
more bonds can be placed. This completely defined the topology of the network.
The surfaces would usually present a problem for this method of forming
connections since at the surface the nearest neighbors are predominantly
`behind' the atom in question. We overcome this difficulty by employing
periodic boundary conditions. For instance, a vertex in the `front' $y-z$
surface, that is, a vertex with $x\geq0$ can connect to vertices in the `back'
$y-z$ surface ($x\geq L$) by connecting forwards. In essence, instead of there
being a single network, there is a central network (in which we form
connections), but this network is surrounded by 26 `image networks' which
connect to the 6 faces, 12 edges and 8 corners such that connections from the
central network can connect to a vertex in any one of the 26 surrounding
networks if they happen to be one of the z nearest neighbors. In this way
connections are not biased at the surfaces.  Once the topology of the network
is defined in this way, we deform the network affinely by a very small shear
strain with  typically $\epsilon =10^{-4}$ (or $\epsilon =0.01\%$). After the
affine deformation we define a non-affine box, which is centered on the centre
of the original network but extends out to 90\% of the distance in the x,y,z
directions compared to the original network. Each bond on a vertex exerts a
force according to Hooke's law. If a vertex lies inside the non-affine box, the
position is permitted to move under the influence of the force in an
over-damped way (the time-step update is such that the positions are altered in
the direction of, and with a magnitude proportional to, the force). All
vertices outside the non-affine box, are not permitted to move and must
therefore deform affinely. It is via these surface vertices that the strain
constraint is implemented. This relaxation procedure is repeated 104 times and
the energy of the network inside the non-affine box is recorded as $E^{NA}$.
Each such relaxation procedure is repeated multiple times for each value of
disorder. The shear modulus of the network is then inferred from this using the
formula: $G = 2E^{NA}/\epsilon^2$, where we have assumed that the contributions
to the modulus from the terms on the order of $\epsilon^4$ can be neglected at
deformations this small.

Many of theoretical results in this paper hold in the limit of large system
size. To verify that the effects observed in our simulations are not merely
finite size effects we repeated the simulations over a wide range of system
sizes  ($N_{min}=64$, $N_{max}=8000$), and observed that, as the system size
increases, the curves of modulus against disorder approach saturation in a
smooth manner, which we assume is the limit of infinite system size. Figure 3a
in the main text illustrates this convergence.

The method of relaxation used in our simulations necessarily means that one
will never truly reach equilibrium in a finite number of time-steps; at best
the energy will approach its true equilibrium value exponentially. To test that
the effects observed in this paper were not an artefact of incomplete
relaxation of local forces, we similarly plotted the curves of modulus against
disorder for increasing number of relaxation steps (equivalent to the time
allowed for stress relaxation), ranging from $10$ up to $10^5$ in powers of
$10^{1/2}$. Figure 3b in the main text illustrates this convergence. As in the
case of finite size effects, the effect of finite relaxation time is that upon
increasing the total number of steps, the curves converge smoothly to a limit
which we take as the infinite-time one. We are therefore confident that neither
finite size, nor finite time effects are responsible for the observed
stiffening of network with disorder, but that the effect is an inherent
property of the network.


\begin{thebibliography}{99}

\bibitem{alexander} Alexander, S. Amorphous solids: their structure,lattice
dynamics and elasticity. Phys. Rep. 296, 65 (1998).
\bibitem{lubensky} DiDonna, B. A. and Lubensky, T. C. Nonaffine correlations in
random elastic media. Phys. Rev. E 72, 066619 (2005).
\bibitem{lemaitre} Lemaitre, A. and Maloney, C. Sum Rules for the Quasi-Static
and Visco-Elastic
       Response of Disordered Solids at Zero Temperature. J. Stat. Phys. 123,
415 (2006).
\bibitem{barrat} Tanguy, A. et al. Continuum limit of amorphous elastic bodies:
A finite-size study of low-frequency harmonic vibrations. Phys. Rev. B 66, 17
(2002).
\bibitem{behringer} Utter, B. and Behringer, R.P. Experimental measures of
affine and nonaffine deformation in granular shear. Phys. Rev. Lett. 100,
208302 (2008).
\bibitem{anderson} Anderson,  P.W.      More is different - broken symmetry and
nature of hierarchical structure of science. Science 177, 393 (1972).
\bibitem{anderson_book} Anderson, P.W. in Ill-Condensed Matter, Les Houches
Session XXXI, Eds. Balian, R., Maynard, R., and Toulouse, G. (North-Holland,
Amsterdam, 1979).
\bibitem{vanhecke} van Hecke, M. Jamming of soft particles: geometry,mechanics,
scaling and isostaticity. J. Phys.: Cond. Mat. 22, 033101 (2010).
\bibitem{debenedetti} Debenedetti, P.G. and and Stillinger, F.H. Supercooled
liquids and the glass transition. Nature 410, 259 (2001).
\bibitem{thorpePRB} Feng, S., Thorpe, M. F. and Garboczi, E. Effective-medium
theory of percolation on central-force elastic networks. Phys. Rev. B 31, 276
(1985).
\bibitem{thorpePRL} He, H. and Thorpe, M. F. Elastic properties of glasses.
Phys. Rev. Lett. 54, 2107 (1985).
\bibitem{mackintosh} Das, M., MacKintosh, F.C. and Levine, A.J. Effective
medium theory of semiflexible filamentous networks. Phys. Rev. Lett. 99, 038101
(2007).
\bibitem{frey} C. Heussinger and E. Frey, Phys. Rev. Lett. {\bf 96}, 017802
(2006).
\bibitem{born} Born, M. and Huang, H. The Dynamical Theory of Crystal Lattices,
p.248 (Oxford University Press, Oxford, 1954).
\bibitem{thorpe_phillips} Phillips, J.C. and Thorpe, M.F. Constraint theory,
vector percolation and glass-formation. Sol. State Commun. 53, 699 (1985).
\bibitem{yabuuchi} Yabuuchi, T. et al. New high-pressure phase of calcium. J.
Phys. Soc. Japan 74, 2391 (2005).
\bibitem{djohari} Djohari, H., Milstein, F. and Maroudas, D. Stability of
simple cubic crystals. Appl. Phys. Lett. 90, 161910 (2007).
\bibitem{legut} Legut, D., Friak, M. and Sob, M. Why is polonium simple cubic
and so highly anisotropic? Phys. Rev. Lett. 99, 016402 (2007).
\bibitem{zaccone} Zaccone, A. and Scossa-Romano, E. Approximate analytical
description of the nonaffine response of amorphous solids. Phys. Rev. B 83,
184205 (2011).
\bibitem{makse} Makse, H. A., Gland, N., Johnson, D.L. and Schwartz, L.M. Why
Effective Medium Theory Fails in Granular Materials. Phys. Rev. Lett. 83, 5070
(1999).
\bibitem{ohern} O'Hern, C.S. et al. Jamming at zero temperature and zero
applied stress: The epitome of disorder. Phys. Rev. E 68, 011306 (2003).
\bibitem{lubensky_2010} Mao, X., Xu, N. and Lubensky, T.C. Soft Modes and
Elasticity of Nearly Isostatic Lattices: Randomness and Dissipation. Phys. Rev.
Lett. 104, 085504 (2010).
\bibitem{obukhov} Obukhov, S. First-order rigidity transition in random rod
networks. Phys. Rev. Lett. 74, 4472 (1995).
\bibitem{moukarzel} Moukarzel, C., Duxbury, P.M. and Leath, P.L.
Infinite-cluster geometry in central-force networks. Phys. Rev. Lett. 78, 1480
(1997).
\bibitem{wyart} Wyart, M. On the rigidity of amorphous solids. Ann. Phys.
(Paris) 30, 1 (2005).
\bibitem{nelson} Steinhardt, P.J., Nelson, D.R. and Ronchetti, M.
Bond-orientational order in liquids and glasses. Phys. Rev. B 28, 784 (1983).
\bibitem{debenedetti_book} Debenedetti, P. G. Metastable Liquids (Princeton
University Press, Princeton NJ, 1996).
\bibitem{andersen} Jonsson, H. and Andersen,    H.C. Icosahedral ordering in
the Lennard-Jones liquid and glass. Phys. Rev. Lett. 60, 2295 (1988).
\bibitem{angell} Angell, C.A. Spectroscopy simulation and scattering, and the
medium range order problem in glass. J. Non-Cryst. Solids 73, 1 (1985).
\bibitem{phillips} Phillips, J. C. Topology of covalent non-crystalline solids
2. Medium-range order in chalcogenide alloys and a-Si(Ge) J. Non-Cryst. Solids
43, 37 (1981).
\bibitem{sheng} Sheng, H.W. et al. Atomic packing and short-to-medium-range
order in metallic glasses. Nature 439, 419 (2006).

\end{thebibliography}
\end{document}